\documentstyle[prd,aps]{revtex}
\newcommand{\e}{\epsilon}
\newcommand{\la}{\lambda}
\newcommand{\ka}{\kappa}
\newcommand{\de}{\delta}
\newcommand{\g}{\gamma}
\newcommand{\al}{\alpha}
\newcommand{\be}{\beta}

\begin {document}
\draft

\title {Fermion helicity flip by parity violating torsion}
\author{Soumitra SenGupta \footnote{Electronic address: {\em
soumitra@juphys.ernet.in}} ${}^{(1)}$ and 
Aninda Sinha \footnote{Electronic address: {\em
as402@cam.ac.uk}}${}^{(2)}$}
\address{{\rm$^{(1)}$ Department of Physics,
Jadavpur University, Calcutta 700 032, India}}\address{{\rm$^{(2)}$
St.Edmund's College, University of Cambridge, Cambridge CB3 0BN, UK}}
\maketitle

\begin{abstract}

{The helicity flip of a spin-1/2 Dirac fermion interacting with a torsion-
field endowed with a pseudo-tensorial extension is analysed. Taking
the torsion to be represented by a Kalb-Ramond field,  we show that
there is a finite amplitude for helicity flip
for massive fermions. The lowest order contribution which turns out to be  
proportional to the pseudo-tensor term, implies a new physical
understanding of the phenomenon of helicity flip.\\}
\end{abstract}
\maketitle

In the past few years, research in the field of space-time with torsion has
yielded a number of interesting results. In particular, the extension of
the Dirac-Einstein Lagrangian by incorporating a Cartan term, has been studied
extensively. This is due to the fact that the spin of a particle turns out to be 
related to torsion just as mass is responsible for curvature \cite{hehl}.  The 
possibility for a flip in helicity which is of enormous interest in the context 
of the early universe and solar neutrinos, has been 
explored \cite{capo,aldro,raj}. It has been proved that in the presence of 
torsion, helicity is not conserved \cite{capo}. 

One very important observation is that the presence of torsion destroys the 
cyclic property of the Riemann-Christoffel tensor. As a result, the standard 
Einstein-Hilbert action admits  a parity-violating term \cite{hoj}.
Since the very presence of torsion automatically
allows parity-violation in the Lagrangian for pure gravity, it should be
possible to incorporate the latter in the geometry of space-time itself.
In a previous paper\cite{ssg}, SS in collaboration with Mukhopadhyaya showed that 
this could be achieved by suitably extending the covariant derivative
with a set of pseudo-tensorial connections, proportional to the torsion
tensor itself. This gave a unique prescription in terms of the
covariant derivatives to obtain parity-violating effects in the Lagrangians
for pure gravity as well as for matter fields with arbitrary
spin. Such a theory not only allows one to couple electromagnetic
field with torsion in a gauge invariant way but is also supported by
string theory\cite{pmssg}.

Here we show that a space-time dependent torsion, endowed with a
pseudo-tensorial extension can
produce a flip. The flipping amplitude for our process turns out to be dependent
solely on
the extension and vanishes had the extension not been there. In case of massive 
fermions where a chirality state is
a linear superposition of two helicity states and vice versa, the
helicity flip obviously results in chirality flip as well. 
Throughout the calculation we take a quantized massive fermion field in a
classical Kalb-Ramond background.
The motivation of this work is to find an alternative prescription which may
explain the helicity flip of a massive fermion from a physical standpoint
so far unexplored.   
For a non-vanishing neutrino mass this result may turn out to be significant in
the context of the solar neutrino problem.

The most general connection in space-time geometry includes, in addition
to the symmetric and antisymmetric parts, a pseudo-tensorial part as well.
Such a pseudo-tensorial extension can lead to parity violation in gravity.
It was shown \cite{ssg} that the most general affine connection is:

\begin {equation}
\tilde {\Gamma}^{\ka}_{\nu\la}=\Gamma^{\ka}_{\nu\la}
-H^{\ka}_{\nu\la}-q(\e^{\g\de}_{\nu\la}H^{\ka}_{\g\de}-\e^{\ka\al}_{\be\la}
H^{\be}_{\g\al}+\e^{\ka\al}_{\be\nu}H^{\be}_{\la\al}) \label{pseudo}
\end{equation}
Here $\Gamma$'s are the usual Christoffel symbols, $H$ is the contorsion
tensor, 
q is parameter determining the extent of parity violation, depending, presumably 
on the matter and spin distribution and so can be helicity dependent.
It has been shown \cite{pmssg} that the torsion tensor H can be equated to the field
strength of the background second rank massless antisymmetric tensor field namely the
Kalb-Ramond field which appears in the massless sector of the string multiplet. 

Using (\ref{pseudo}) to extend the Dirac-Einstein Lagrangian for a 
spin-1/2 field\cite{audr,fig} in a spacetime  with torsion, we get:
\begin{equation}
{\mathcal{L}}_{Dirac}=\bar{\psi}[i\g^{\mu}(\partial_{\mu}-
\sigma^{\rho\be}v^{\nu}_{\rho}g_{\la\nu}\partial_{\mu}v^{\la}_{\be}-
g_{\al\de}\sigma^{ab}v^{\be}_{a}v^{\de}_{b} \tilde{\Gamma}^{\al}_{\mu\be})]\psi
\label{L}
\end {equation}
where one has introduced the tetrad $v^{\mu}_{a}$ to connect the curved
space with the corresponding tangent space at any point and  
$\sigma^{ab}$ is the commutator of the Dirac matrices. The Greek
indices correspond to the curved space and the Latin indices to the tangent
space. Using the full form of $\tilde{\Gamma}$ defined in (\ref{pseudo}) $,
{\mathcal{L}}_{Dirac}$ can be expressed as
\begin {equation}
{\mathcal{L}}_{Dirac}={\mathcal{L}}^{E}+\bar {\psi}[i\g^{\mu}g_{\al\de}\sigma^{\be\de}
H^{\al}_{\mu\be}]\psi+{\mathcal{L}}^{pv} \label{br}
\end{equation}

The first term is the one obtained in Einstein gravity, the second corresponds
to the Cartan extension(${\mathcal{L}}^{C}$). The third term results from the 
pseudo-tensorial extension of the affine connection and is responsible for
parity violation. This term is given later explicitly in equation (5).

${\mathcal{L}}_{Dirac}$ is used to calculate the transition amplitude
from one helicity state to another. By helicity states we mean the
spinors corresponding to the eigen-states of the operator $\Sigma
. \hat{p}$. Presumably one could work also with the chirality states,
the spinors corresponding to the eigen-states of $\g_{5}$.  In the presence of torsion,
helicity is not conserved \cite{capo} and as such its flipping is considered. To
calculate the flip we remember that the field H being a classical field can only
appear as the external lines in the Feynman diagrams. 
We therefore consider the following lower order  Feynman diagrams:
\\
\\
\begin{center}
\begin{picture}(150,100)

\thicklines
\put(0,0){\line(1,1){49}}
\put(25,25){\vector(1,1){10}}
\put(50,50){\circle*{5}}
\put(50,50){\vector(1,0){60}}
\put(10,2){$u_s(p_1)$}
\put(112,52){$u_{s'}(p_2)$}
\multiput(0,95)(10,-10){5}{\begin{picture}(2,2) \put(0,0){\oval(10,10)[tr]} \put(10,0){\oval(10,10)[bl]}
\end{picture}}
\put(10,100){$H(k)$}
\put(79,10){(a)}
\end{picture}
\end{center}
\begin{center}
\begin{picture}(150,100)

\thicklines
\put(0,0){\line(1,1){49}}
\put(25,25){\vector(1,1){10}}
\put(50,50){\circle*{5}}
\put(50,50){\vector(1,0){30}}
\put(80,50){\line(1,0){30}}
\put(10,2){$u_s(p_1)$}

\multiput(0,95)(10,-10){5}{\begin{picture}(2,2) \put(0,0){\oval(10,10)[tr]} \put(10,0){\oval(10,10)[bl]}

\end{picture}}
\put(0,62){$H(k_{1})$}
\put(110,50){\circle*{5}}
\put(112,50){\line(1,1){49}}
\multiput(118,45)(10,-10){5}{\begin{picture}(2,2)
\put(0,0){\oval(10,10)[bl]} \put(-10,0){\oval(10,10)[tr]}\end{picture}}

\put(137,75){\vector(1,1){10}}
\put(145,62){$u_{s'}(p_{2})$}
\put(165,2){$H(k_{2})$}

\end{picture}
\end{center}

\begin{center}
\begin{picture}(100,160)

\thicklines
\put(100,0
){\line(-1,1){49}}
\put(80,20){\vector(1,-1){10}}
\put(50,50){\circle*{5}}

\put(50,50){\line(0,1){60}}
 \put(2,120){$u_s(p_1)$} 
\multiput(2,0)(10,10){5}{\begin{picture}(2,2) \put(0,0){\oval(10,10)[tl]} \put(0,10){\oval(10,10)[br]} 

\end{picture}} \put(10,2){$H(k_{1})$}
\put(50,110){\circle*{5}}
\put(50,112){\line(-1,1){49}}
\multiput(55,110)(10,10){5}{\begin{picture}(2,2) \put(0,0){\oval(10,10)[tl]} \put(0,10){\oval(10,10)[br]} \end{picture}}


\put(25,137){\vector(1,-1){10}}
\put(60,2){$u_{s'}(p_{2})$}
\put(75,120){$H(k_{2})$}

\end{picture}
\end{center}

\begin{center}
(b)\\
t $\rightarrow$\\ 
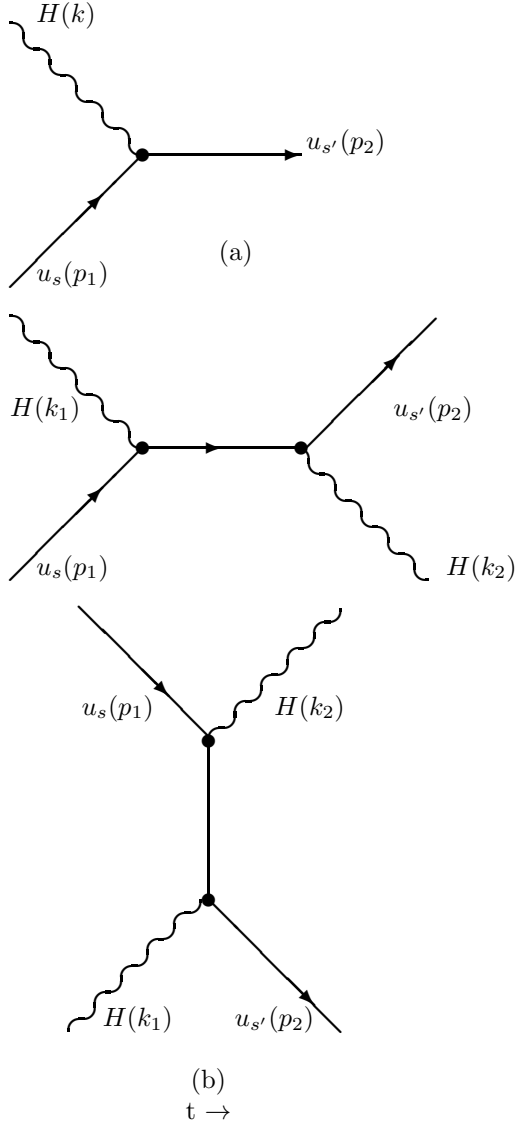
Figure 1: Leading order Feynman diagrams for helicity flip
\end{center}
For a weak gravitational field, 
we set $g_{\al\be}=\eta_{\al\be}$ where $\eta_{\al\be}$ is the
Minkowski metric diag(1,-1,-1,-1).   H cannot be a constant since it
implies that the field H carries no energy or momentum, which cannot be
possible. As was argued in \cite{pmssg}, H can be written as the Hodge-dual to the 
derivative of a massless pseudo-scalar field $\tilde{\phi}$, which is the axion of
the corresponding String theory.We then get:

\begin{equation}
{\mathcal{L}}^{C}=i\bar {\psi} \g^{\mu} \sigma^{\be\phi} \epsilon_{\phi\mu\be\la}( \partial^{\la}\tilde{ \phi}) \psi  \label{cartan}
\end{equation}

\begin{equation}
{\mathcal{L}}^{pv}=q \bar {\psi} \g^{\al} \sigma^{\g \nu} \g_{5} \epsilon_{\al\g\nu\la}
(\partial^{\la}\tilde{\phi}) \psi
\label{pv}
\end{equation}

${\mathcal{L}^E}$ doesn't contribute in flat space. As such,
the calculations are done using ${\mathcal{L}}_{int}
={\mathcal{L}}^{C}+{\mathcal{L}}^{pv}$.\\

It may be noted that both the parity violating and parity conserving interaction
Lagrangians are dimension five operators leading to the usual problem of
renormalizability in a full quantum theory. However, here we are considering a
semi-classical theory and therefore have the leading order finite amplitude diagrams
as drawn above.

Now we first consider that the field H is spatially homogeneous
and depends only on time. For an isotropic and homogeneous
background geometry this is a valid assumption.

With such a choice for H, 
if we now consider the Feynman diagram of figure 1(a), we get a factor
$\delta^4(p_1-p_2+k)=\delta^3({\bf{p_1}-\bf{p_2}+\bf{k}})\delta(p^{0}_{1}-p^{0}_{2}
+k^{0})$
in the S matrix element. This simplifies to a proportionality to
$\delta(k^{0})$ where $k^{0}$ corresponds to the energy of the H field. This
will
contribute only if $k^{0}$ is zero. However since there is only a t-dependence,
$k^{0}$ 
cannot be zero since in that case E=0 which means that the field H vanishes
identically. 
Since 1(a) is the fundamental vertex, it is clear that there cannot be a helicity 
flip. As such a pure time-dependence of H {\it{cannot}} produce a helicity flip.

Next we consider that H depends on both space and time.
A similar argument as above shows that 1(a) contributes and as such there can be 
a helicity flip for such an H. 
The transition-matrix element between $s$-helicity state to $s'$-helicity
state, in accordance with the Feynman diagram of figure 1(a), is:
\begin {equation}
{\cal{M}}_{if}=6 \left[\bar {u}_{s'}(q+\g_{5})\not{k} u_s \right] 
\end {equation}
where $\not{k}=\g^{\mu} k_{\mu}$.

We use the gamma matrices and spinors in the Dirac-Pauli
representation \cite{hal}.

Assuming the fermion to be initially moving in the z-direction, the result for $+\frac{1}{2}$ helicity to $-\frac{1}{2}$ helicity or vice versa is:
\begin{equation}
{\cal{M}}_{if}=12k_x m \sqrt{\frac{E+m}{E'+m}}
\end{equation}
where $E'=E+k^0$.

Note that this amplitude is independent of the pseudo-tensor term.
It also suggests that the field H has to have
a non-zero momentum component in a direction that is different from
the fermion direction. In fact, if in the lab frame, the momentum of the field H
points in the same direction as that of the fermion, there cannot be a
helicity flip. The transition amplitude is proportional to mass, but
it is modified by the presence of $k_x$. However this cannot be the
lowest order contribution since it is kinematically forbidden. This can be seen by noting that the energy-momentum relations and the mass constraints cannot be satisfied simultaneously.

Now let's consider the next order Feynman diagrams in fig 1b.
The second diagram in 1b is  analogous to
the crossed diagram that is obtained for Compton scattering.  
The transition-matrix element between $s$-helicity state to $s'$-helicity
state, in accordance with the Feynman diagram of figure 1(b), is:
\begin {equation}
{\cal{M}}_{if}= \left[36 \bar {u}_{s'} (q'+\g_{5})\not{k_{2}}
\frac{\not{p_{1}}+\not{k_{1}} +m}
{(p_{1}+k_{1})^2-m^2}(q+\g_{5})\not{k_{1}} u_s \right] \label{M}
\end {equation} 

For the 2nd diagram in figure 1(b), we get:
\begin {equation}
{\cal{M}}_{if}= \left[36 \bar {u}_{s'} (q'+\g_{5})\not{k_{1}}
\frac{\not{p_{1}}-\not{k_{2}} +m}
{(p_{1}-k_{2})^2-m^2}(q+\g_{5})\not{k_{2}} u_s \right] \label{M'}
\end {equation} 

We use $q'=-q$. In doing so, we assume that $q$ is
helicity-dependent. The inspiration behind this is as follows. In the
matrix element (\ref{M}), there are terms like $(q+\gamma_{5})$. It is
well known that for highly energetic fermions, $(1+\gamma_{5})$ seeks
out the right-handed particles while $(1-\gamma_{5})$
seeks out the
left-handed particles. In order to get terms
resembling these, it is
necessary to make $q$ helicity-dependent. If we don't make such a
choice we get other transitions which are briefly mentioned later.

Let's consider eqn(\ref{M}).
Put  $p_{1}+k_{1}=K$.
From here, the calculation can be simplified. Writing
$\not{k_{1}}=k_{1} \cdot \gamma$ and applying the Dirac equation, we
get
\begin{equation}
\not{k_{1}}u_{s}=(\gamma .K-m)u_{s}
                \end{equation}
Similarly,
\begin{equation}
{\bar{u}}_{s'}\not{k_{2}}={\bar{u}}_{s'}(\gamma.K-m)
\end{equation}
Using,
\begin{equation}
(\gamma .K-m)(\gamma .K+m)=K^{2}+m^{2}
\end{equation}

and analogous relations for the 2nd diagram in figure 1(b),
it is now easy to verify that the  Feynman
diagrams of fig 1b  generate the following result:

\begin{equation}
{\cal{M}}_{if}=144mq \bar{u}_{s'}(p_{2})\gamma_{5} u_{s}(p_{1}) \label{final}
\end{equation}

This shows that the Einstein-Cartan contribution to the matrix element
vanishes and we are left with a contribution proportional to
$q$. {\it{Thus only the pseudo-tensor extension can produce a
flipping.}} Also the flipping amplitude is proportional to
mass. Curiously enough, the only dependence on the H field is in
the factor $q$.

We evaluate (\ref{final}) for a particular case. Suppose the
interaction is confined in the x-z plane. Let the z-axis be aligned
with the initial direction of the fermion before it interacts. Using
the Dirac-Pauli representation as in \cite{hal}, the following result
is obtained:

\begin{equation}
{\cal{M}}_{if}=144qm p_{2x} \sqrt{\frac{E+m}{E_{2}+m}}
\end{equation}

Here subscript 2 refers to the fermion after interaction. The
amplitude is for right-handed helicity to left-handed helicity
transition or vice-versa. Obviously the flipping amplitude vanishes for zero-mass fermions.

We have used a Dirac-Einstein Lagrangian modified by torsion and  a
pseudo-tensorial extension, to calculate the flipping between helicity
states for massive fermions. It was shown that for a specific process, the flipping
probability is non-zero for  massive fermions and solely depends
on the extension. The crucial assumption that we have made is that the
coupling of the extension term can be helicity dependent. Presumably
it behaves something like the electric charge and is equal and
opposite for opposite helicities. This assumption makes the Feynman
rules for our process resemble the V-A rules in Weak Interaction.
It is important to note that while one gets a nonvanishing flipping 
probability from the fermion mass itself, here we find an additional 
contribution from the pseudo tensorial extension of the torsion. This 
extra term contains the coupling q. 

It is interesting to note that even if q were to be independent of helicity, we get a
finite amplitude for helicity flipping for particle to anti-particle
transitions. Such transitions are probably allowed for our Lagrangian
since it doesn't conserve ${\bf{C}}$. Such analysis can form the basis
of future ventures in this direction.

Our work thus opens up a completely new physical possibility, namely parity
violating torsion, to explain the spin flip of massive fermions. Whether the
fermions couple to the background in the way discussed here, depends on the type
of theoretical model we consider. However in a string theory inspired model where
we indeed have a Kalb-Ramond background, such a coupling occurs naturally.
In this context we propose that the result of this work can be a possible
explanation for the so called solar neutrino problem where the
number of left handed neutrino detected is less than that expected. 
For a quantitative estimation one needs to know the exact nature of the torsion
field and its parity violating extension inside the sun. 
Although we already have a bound on the classical axion, the undetermined
coupling parameter 'q' appearing in our theory may be determined from known
experimental results.
As the contribution of the helicity flip from the neutrino mass, which is very small,
fails to explain  the observed shortage of left handed neutrino, this work proposes 
an additional effect in terms of the torsion coupling `q' as a possible 
explanation to this longstanding problem. Work in this direction is in progress
and will be reported elsewhere.

\section*{\small\bf{Acknowledgements}} 
We acknowledge stimulating discussion with A. Chatterjee, T. De,
R. Gandhi, B. Mukhopadhyay and A. Raychaudury.
S.S. is supported by project grant no. 98/37/16/BRNS cell/676 from the
Board Of Research In Nuclear Sciences, Department Of Atomic Energy, Government Of
India. A.S. is supported by a scholarship from Cambridge Commonwealth Trust.

\end{document}